\documentclass[reprint,aps,prd,twocolumn,notitlepage,nofootinbib,showpacs,preprintnumbers,groupedaddress]{revtex4-1}
\usepackage[T1]{fontenc}
\usepackage[utf8]{inputenc}
\usepackage{bm}
\usepackage{amsmath}
\usepackage{amssymb}
\usepackage{esint}
\usepackage{color}
\usepackage{graphicx}
\PassOptionsToPackage{normalem}{ulem}
\usepackage{ulem}

\usepackage{times}
\usepackage{hyperref}

\begin{document}

\title{Spatially covariant gravity theories with two tensorial degrees of freedom: the formalism}

\author{Xian Gao}%
\email[Email: ]{gaoxian@mail.sysu.edu.cn}

\author{Zhi-Bang Yao}%
\email[Email: ]{yaozhb@mail2.sysu.edu.cn}

    \affiliation{%
        School of Physics and Astronomy, Sun Yat-sen University, Guangzhou 510275, China}

\date{October 31, 2019}

\begin{abstract}
	Within the general framework of spatially covariant theories of gravity, we study the conditions for having only the two tensorial degrees of freedom. Generally, there are three degrees of freedom propagating in the theory, of which two are tensorial and one is of the scalar type. Through a detailed Hamiltonian analysis, we find two necessary and sufficient conditions to evade the scalar type degree of freedom. The first condition implies that the lapse-extrinsic curvature sector must be degenerate. The second condition ensures that the dimension of the phase space at each spacetime point is even, so that the scalar type degree of freedom is eliminated completely. We also compare our results with the previous studies, and apply our formalism to a simple example, in which the Lagrangian is quadratic in the extrinsic curvature.
\end{abstract}

\maketitle

\section{Introduction}

Although has passed all the ground and space tests in the past one hundred years, Einstein's General Relativity (GR) is still being questioned on its uniqueness and if it is the true theory of gravity of Nature.
This question can be made more concrete after the detection of gravitational waves \cite{Abbott:2016blz}, which are the two tensorial degrees of freedom (DoFs) in the spatially homogeneous and isotropic background. That is, if GR is the unique theory that propagates only the two tensorial DoFs?

This question has partially been answered by Lovelock \cite{Lovelock:1971yv,Lovelock:1972vz}, who has proved that GR is the unique theory for the metric with second derivatives, which preserves spacetime diffeomorphism and locality in the four dimensional spacetime.
As a result, the question may be asked alternatively: how to encode the two tensorial DoFs consistently in a field theory that is different from GR, without introducing other DoFs and pathologies?

One kind of theories we may use to encode the two tensorial DoFs are the generally covariant scalar-tensor theories, which in general include higher order derivatives of both the scalar field and the metric.
In the past decades, much effort has been made to keep the number of DoFs no larger than 3, by evading the extra DoFs arising due to the presence of higher order derivatives \cite{Horndeski:1974wa,Deffayet:2011gz,Kobayashi:2011nu,Gleyzes:2014dya,Gleyzes:2014qga,Langlois:2015cwa,Motohashi:2016ftl} (see Refs. \cite{Langlois:2018dxi,Kobayashi:2019hrl} for reviews).
For our purpose, the task is in some sense more aggressive, that is we wish to eliminate the scalar DoFs completely, although the theory is originally parametrized with a scalar field as one of the variables of its configuration space.

Some progresses have been made along this direction within the framework of ``cuscuton'' theory \cite{Afshordi:2006ad,Afshordi:2007yx} and its generalizations \cite{Gomes:2017tzd,Iyonaga:2018vnu}.
Generally, the cuscuton theory is defined to be scalar-tensor theories that propagate only two tensorial DoFs in the unitary gauge, that is when the scalar field is chosen to be spatially homogeneous and isotropic.
The cuscuton theory partially solved the task to encode the two tensorial DoFs in the framework of scalar-tensor theories, in the sense that the gradient of the scalar field has to be timelike.
In this sense, the cuscuton theory can be viewed as a Lorentz breaking, and in particular, spatially covariant gravity theory \cite{Afshordi:2009tt,Bhattacharyya:2016mah}.

Indeed, when fixing the so-called unitary gauge with $\phi = t$, the generally covariant scalar-tensor theory can always be written in terms of spatially covariant gravity theories, in which the basic building blocks are spatially covariant tensors, such as the spatial metric $h_{ij}$, extrinsic curvature $K_{ij}$ and intrinsic curvature $R_{ij}$ as well as their spatial and temporal derivatives.
Spatially covariant theories of gravity have been studied previously with different motivations and forms, such as in the effective field theory of inflation \cite{Creminelli:2006xe,Cheung:2007st}, in the Ho\v{r}ava gravity \cite{Horava:2009uw,Blas:2009qj}, etc.
Spatially covariant gravity theories with at most three degrees of freedom were systematically studied in \cite{Gao:2014soa,Gao:2014fra,Fujita:2015ymn,Gao:2018znj,Gao:2019lpz,Gao:2018izs,Gao:2019liu}.
When the general covariance is apparently recovered, spatially covariant gravity theories can also be used as ``generators'' of healthy scalar-tensor theories with higher derivatives \cite{Gao:toappear}.
Generally, working in the unitary gauge may be risky, since the mode that disappears in the unitary gauge would re-arises when apparently recovering the general covariance. Nevertheless, it was argued in \cite{DeFelice:2018mkq} that such an extra mode is superficial and can be safe by choosing appropriate boundary conditions.

It is thus more convenient to work directly in the framework spatially covariant gravity theories to study how to build a theory with only two tensorial Dofs.
Generally, the spatially covariant gravity theories with only spatial derivatives propagate 3 local DoFs. Thus one needs to impose further constraints on the form of the theory in order to reduce the number of DoFs, i.e., to eliminate the scalar type DoF.
A class of ``minimally modified gravity'' was proposed in \cite{Lin:2017oow,Aoki:2018brq} (see also \cite{Carballo-Rubio:2018czn}), which studied the Lagrangian that is linear in the lapse function and derived the conditions for having two tensorial DoFs at the level of Lagrangian. The conditions were also studied directly at the level of the Hamiltonian \cite{Mukohyama:2019unx}.

In light of these progresses, it is interesting to study the conditions for having two tensorial DoFs in the general framework of spatially covariant gravity theories, in particular, at the level of Lagrangian. In this work, we take a first step and start from the framework proposed in \cite{Gao:2014soa,Gao:2014fra}, in which the Lagrangian is a general function of $N$, $h_{ij}$, $R_{ij}$ and their spatial derivatives.
By performing a Hamiltonian analysis, we will derive under which conditions the scalar DoF can be removed completely.

This rest of the paper is organized as following. 
In Sec. \ref{sec:Non-perturbative-Hamiltoina-anal}, we setup our formalism for the Hamiltonian analysis, following the formalism developed in \cite{Gao:2018znj}.
In Sec. \ref{sec:TTcond} we derive the conditions for having two tensorial DoFs. In particular, two conditions are needed, which we dub as the first and the second TTDoF conditions, respectively.
In Sec. \ref{sec:app}, we will compare our results with the previous studies and illustrate our formalism with a Lagrangian that is quadratic in the extrinsic curvature.
Sec. \ref{sec:Conclusion} is for conclusion.

\section{Hamiltonian analysis} \label{sec:Non-perturbative-Hamiltoina-anal}

In this section, we perform the Hamiltonian analysis to derive the conditions of the theory in order to have two tensorial degrees of freedom. We will follow the formalism developed in \cite{Gao:2018znj}.

\subsection{The action}

The spatially covariant gravity is constructed based on the foliation structure of the spacetime. The basic ingredients of the Lagrangian are spatially covariant tensorial quantities, i.e., the normal vector $n_{\mu}$ and the induced metric $h_{\mu\nu}$ of the spacelike hypersurfaces. Due to the foliation structure, there are two types of derivatives, the spacelike one $\mathrm{D}_{\mu}$, which is the covariant derivative compatible with the induced metric, and the timelike one $\pounds_{\vec{n}}$, which is the Lie derivative with respect to the normal vector.

The general action for the spatially covariant gravity is thus
	\begin{equation}
	S=\int\mathrm{d}^{4}x\,\sqrt{-g}\,\mathcal{L}\left(\phi,N,h_{\mu\nu},{}^{3}\!R_{\mu\nu},\mathrm{D}_{\mu},\pounds_{\vec{n}}\right),
	\end{equation}
where $\phi$ is the scalar field which specifies the hypersurfaces (that is, each hypersurface is defined by $\phi=\mathrm{const.}$), and $N$ is defined through $n_{\mu} = -N \nabla_{\mu}\phi$.
Since the hypersurfaces are spacelike, we are allowed to choose a coordinate system compatible with the foliation structure, which is nothing but the ADM coordinates, in which the action can be written as
	\begin{equation}
		S = \int \mathrm{d}t\mathrm{d}^3 x \,N\sqrt{h} \mathcal{L}\left( t,N,h_{ij},R_{ij},\nabla_{i},\pounds_{\vec{n}}\right).
	\end{equation}
Such a choice of coordinates is also dubbed as the unitary gauge in the literature.
Throughout this paper, $\nabla_{i}$ is understood as the covariant derivative compatible with $h_{ij}$.
Generally, $N$ and $h_{ij}$ are independent and thus both acquire ``time'' derivatives through $\pounds_{\vec{n}}N$, $\pounds_{\vec{n}}h_{ij}$, etc.
This case was discussed in \cite{Gao:2018znj}.
In this paper, as a first step, we concentrate on case with only $\pounds_{\vec{n}}h_{ij} \equiv 2 K_{ij}$.

We thus start from the action \cite{Gao:2014soa,Gao:2014fra}
	\begin{equation}
	S=\int \mathrm{d}t\mathrm{d}^{3}x\,N\sqrt{h}\,\mathcal{L}\left(N,h_{ij},K_{ij},R_{ij},\nabla_{i};t\right),\label{S_SCG I}
	\end{equation}
where 
	\begin{equation}
	K_{ij}\equiv \frac{1}{2} \pounds_{\vec{n}}h_{ij}=\frac{1}{2N}\left(\dot{h}_{ij}-\pounds_{\vec{N}}h_{ij}\right),
	\end{equation}
is the extrinsic curvature with $\pounds_{\vec{N}}$ denoting the
Lie derivative with respect to the shift-vector $N^{i}$, $R_{ij}$ is the Ricci tensor of $h_{ij}$. In (\ref{S_SCG I}), $\nabla_{i}$ is allowed to act on other fields with arbitrary orders.

The nonlinear dependence on the ``velocity'' $\dot{h}_{ij}$ of the Lagrangian makes the explicit reversion of velocity in terms of the momentum impossible. 
Following the same strategy in \cite{Gao:2018znj} (see also \cite{Saitou:2016lvb}), we use an equivalent action that is linear in the velocity by introducing an auxiliary field $B_{ij}$ and the Lagrange multiplier $\Lambda^{ij}$ as, and write 
	\begin{equation}
	S=S_{B}+\int \mathrm{d}t\mathrm{d}^{3}x\,N\sqrt{h}\,\Lambda^{ij}\left(K_{ij}-B_{ij}\right),\label{S_B+B-K}
	\end{equation}
where the non-dynamical part $S_{B}$ is obtained by simply replacing $K_{ij}$ as $B_{ij}$ in the action (\ref{S_SCG I}), i.e.,
	\begin{equation}
	S_{B}\equiv\int \mathrm{d}t\mathrm{d}^{3}x\,N\sqrt{h}\,\mathcal{L}_{B}\left(N,h_{ij},B_{ij},R_{ij},\nabla_{i};t\right).\label{S_B}
	\end{equation}
The equation of motion for $\Lambda^{ij}$ enforces $B_{ij}$ to be identical with $K_{ij}$, and thus the two actions (\ref{S_SCG I}) and (\ref{S_B+B-K}) are equivalent, at least classically. 
It is also convenient to fix the Lagrange multiplier $\Lambda^{ij}$ by making use of the equation of motion for the auxiliary field $B_{ij}$, which yields
	\begin{equation}
	\Lambda^{ij}=\frac{1}{N\sqrt{h}}\frac{\delta S_{B}}{\delta B_{ij}}.\label{Lagrange multipiler}
	\end{equation}
After these preliminaries, the original
action (\ref{S_SCG I}) can be recast to the  form
	\begin{equation}
	S=S_{B}+\int \mathrm{d}t\mathrm{d}^{3}x\frac{\delta S_{B}}{\delta B_{ij}}\left(K_{ij}-B_{ij}\right).\label{eq:S-1}
	\end{equation}
The action (\ref{eq:S-1}) will be our starting point for the Hamiltonian analysis.

\subsection{The Hamiltonian}

The variables in the action (\ref{eq:S-1}) are $N$, $N^{i}$, $h_{ij}$ and $B_{ij}$.
The corresponding conjugate momenta are
	\begin{equation}
	\pi^{ij}:=\frac{\delta S}{\delta\dot{h}_{ij}}=\frac{1}{2N}\frac{\delta S_{B}}{\delta B_{ij}},\label{pi^ij}
	\end{equation}
	\begin{equation}
	\pi:=\frac{\delta S}{\delta\dot{N}}=0,\qquad\pi_{i}:=\frac{\delta S}{\delta\dot{N}^{i}}=0,\qquad p^{ij}:=\frac{\delta S}{\delta\dot{B}_{ij}}=0.\label{momenta}
	\end{equation}
None of the velocities can be solved from
the relations (\ref{pi^ij}) and (\ref{momenta}). As a result, there are in total 16 primary constraints in our theory:
	\begin{equation}
	\tilde{\pi}^{ij}:=\pi^{ij}-\frac{1}{2N}\frac{\delta S_{B}}{\delta B_{ij}}\approx 0,\label{eq:pi_h tilde}
	\end{equation}
	\begin{equation}
	\pi\approx 0,\qquad\pi_{i}\approx 0,\qquad p^{ij}\approx 0,\label{primary cstr}
	\end{equation}
where and throughout this work ``$\approx$'' represents ``weak equality'' that holds only on the subspace $\Gamma_{P}$ of the phase space defined by the the primary constraints.

For later convenience, we denote the set of variables as
	\begin{equation}
	\left\{ \Phi_{I}\right\} :=\left\{ N^i,h_{ij},N,B_{ij}\right\} ,
	\end{equation}
the set of the conjugate momenta as
	\begin{equation}
	\left\{ \Pi^{I}\right\} :=\left\{ \pi_{i},\pi^{ij},\pi,p^{ij}\right\} ,
	\end{equation}
and the set of primary constraints as 
	\begin{equation}
	\left\{ \varphi^{I}\right\} :=\left\{ \pi_{i},\tilde{\pi}^{ij},\pi,p^{ij}\right\} ,
	\end{equation}
where the indices $I,J,\cdots$ formally denote different kinds of variables as well as their tensorial indices. 
We will also use the indices $a,b,\cdots$ when neglecting the $\left\{ N^{i},\pi_{i}\right\} $-sector, that is 
	\begin{equation}
	\left\{ \varphi^{a}\right\} :=\left\{ \pi,\tilde{\pi}^{ij},p^{ij}\right\} .
	\end{equation}
All kinds of the indices obey the Einstein summation convention.

The canonical Hamiltonian in the subspace $\Gamma_{P}$ is defined by performing the Legendre transformation
	\begin{eqnarray}
	H_{\mathrm{C}}\approx H_{\mathrm{C}}|_{\Gamma} & = & \int \mathrm{d}^{3}x\left(\text{\ensuremath{\Pi}}^{I}\dot{\Phi}_{I}-N\sqrt{h}\mathcal{L}\right)\nonumber \\
	 & \simeq & \int\mathrm{d}^{3}x\left(NC\right)+X[\vec{N}],
	\end{eqnarray}
where 
	\begin{equation}
	C\equiv2\pi^{ij}B_{ij}-\sqrt{h}\mathcal{L}_{B},
	\end{equation}
and for a general spatial vector $\vec{\xi}$, we define a functional $X[\vec{\xi}]$ as \cite{Gao:2018znj}
	\begin{equation}
	X[\vec{\xi}]:=\int\mathrm{d}^{3}x\,\Pi^{I}\pounds_{\vec{\xi}}\,\Phi_{I},\label{Xfn_xi_def}
	\end{equation}
that is
	\begin{equation}
	X[\vec{\xi}]\equiv\int \mathrm{d}^{3}x\Big(\pi_{i}\pounds_{\vec{\xi}}\,N^{i}+\pi\pounds_{\vec{\xi}}\,N+\pi^{ij}\pounds_{\vec{\xi}}\,h_{ij}+p^{ij}\pounds_{\vec{\xi}}B_{ij}\Big).\label{Xfn_xi_xpl}
	\end{equation}	
Here $\vec{\xi}$ may or may not depend on the phase space variables.
One of the advantages of introducing $X[\vec{\xi}]$ is, for an arbitrary functional $\mathcal{F}$ on the phase space that is invariant under the time-independent spatial diffeomorphism, we have the following equality
	\begin{equation}
	\left[X[\vec{\xi}],\mathcal{F}\right]= X\left[[\vec{\xi},\mathcal{F}]\right],\label{PB_Xfn_com}
	\end{equation}
up to a boundary term. We refer to \cite{Gao:2018znj} for the proof for a more general statement.
In (\ref{PB_Xfn_com}) the Poisson bracket $\left[\mathcal{F},\mathcal{G}\right]$ is defined by
	\begin{equation}
		\left[\mathcal{F},\mathcal{G}\right]:=\int\mathrm{d}^{3}z\left(\frac{\delta\mathcal{F}}{\delta\Phi_{I}(\vec{z})}\frac{\delta\mathcal{G}}{\delta\Pi^{I}(\vec{z})}-\frac{\delta\mathcal{F}}{\delta\Pi^{I}(\vec{z})}\frac{\delta\mathcal{G}}{\delta\Phi_{I}(\vec{z})}\right).
	\end{equation}
$X[\vec{\xi}]$ can be recast into the more familiar form through integration by parts
	\begin{equation}
	X[\vec{\xi}]\simeq\int\mathrm{d}^{3}x\,\xi^{i}{C}_{i},\label{Xf_xi_ibp}
	\end{equation}
	with
	\begin{eqnarray}
	{C}_{i} & = & \pi\nabla_{i}N-2\sqrt{h}\nabla_{j}\left(\frac{\pi_{i}^{j}}{\sqrt{h}}\right)\nonumber \\
	&  & +p^{kl}\nabla_{i}B_{kl}-2\sqrt{h}\nabla_{j}\left(\frac{p^{jk}}{\sqrt{h}}B_{ik}\right)\nonumber \\
	&  & +\pi_{j}\nabla_{i}N^{j}+\sqrt{h}\nabla_{j}\left(\frac{\pi_{i}}{\sqrt{h}}N^{j}\right).\label{calCi_def}
	\end{eqnarray}
As we shall see, $C_{i}$ is the generalization of the momentum constraint in GR. 
One also finds the significant properties of $C_{i}$:
	\begin{equation}
	\left[C_{i}\left(\vec{x}\right),Q(\vec{y})\right]\approx 0,\qquad \text{for any} \quad Q\approx 0 ,\label{C_i property}
	\end{equation}
and 
	\begin{equation}
		\left[C_{i}\left(\vec{x}\right),H_{\mathrm{C}}\right]=0, \label{PB_Ci_HC}
	\end{equation}
which simply follow the equality (\ref{PB_Xfn_com}). See also \cite{Gao:2018znj} for the details.

When dealing with the constrained system, the time evolution should be determined by the total Hamiltonian, which is defined as
	\begin{eqnarray}
	H_{\mathrm{T}} & := & H_{\mathrm{C}}+\int d^{3}x\left(\lambda\pi+\lambda^{i}\pi_{i}+\lambda_{ij}\tilde{\pi}^{ij}+\mu_{ij}p^{ij}\right),
	\end{eqnarray}
where $\left\{ \lambda_{I}\right\} \equiv\left\{ \lambda,\lambda^{i},\lambda_{ij},\mu_{ij}\right\} $ are the undetermined Lagrange multiplies. 
In terms of the total Hamiltonian,
the time evolution of any function $Q$ of the phase space variables is given by
	\begin{equation}
	\frac{\mathrm{d}Q}{\mathrm{d}t}\approx\frac{\partial Q}{\partial t}+\left[Q,H_{\mathrm{T}}\right].\label{Q/t}
	\end{equation}
From now on, we assume the Lagrangian $\mathcal{L}$ in (\ref{S_SCG I}) does not depend on $t$ explicitly, i.e., $\partial\mathcal{L}/\partial t \equiv0$. 
It is straightforward to generalize to the explicitly time-dependent cases.

\subsection{The consistency conditions for primary constraints} \label{subsec:The-consistency-condition}

To be consistent, constraints must be preserved in time evolution, otherwise there must exist further constraints. Since we assume $\partial\mathcal{L}/\partial t \equiv0$, all the primary constraints do not depend on time as well. 
The time evolution of the primary constraints (\ref{eq:pi_h tilde}) and (\ref{primary cstr}), i.e. the consistency conditions for the primary constraints, yield
	\begin{equation}
	\int d^{3}y\left[\varphi^{I}\left(\vec{x}\right),\varphi^{J}\left(\vec{y}\right)\right]\lambda_{J}\left(\vec{y}\right)+\left[\varphi^{I}\left(\vec{x}\right),H_{\mathrm{C}}\right]\approx 0. \label{cc_pri}
	\end{equation}
The Poisson brackets involving the primary constraints we evaluated explicitly in \cite{Gao:2018znj}. 
Thus (\ref{cc_pri}) can be written in the matrix form
	\begin{widetext}
	\begin{equation}
	\int \mathrm{d}^{3}y\left(\begin{array}{cccc}
	0 & 0 & 0 & 0\\
	0 & 0 & 0 & [p^{ij}(\vec{x}),\tilde{\pi}^{kl}(\vec{y})]\\
	0 & 0 & 0 & [\pi(\vec{x}),\tilde{\pi}^{kl}(\vec{y})]\\
	0 & [\tilde{\pi}^{ij}(\vec{x}),p^{kl}(\vec{y})] & [\tilde{\pi}^{ij}(\vec{x}),\pi(\vec{y})] & [\tilde{\pi}^{ij}(\vec{x}),\tilde{\pi}^{kl}(\vec{y})]
	\end{array}\right)\left(\begin{array}{c}
	\lambda^{k}(\vec{y})\\
	\mu_{kl}(\vec{y})\\
	\lambda(\vec{y})\\
	\lambda_{kl}(\vec{y})
	\end{array}\right)\approx\left(\begin{array}{c}
	C_{i}(\vec{x})\\
	0^{ij}\\
	C^{\prime}(\vec{x})\\{}
	[H_{\mathrm{C}},\tilde{\pi}^{ij}(\vec{x})]
	\end{array}\right),\label{eq:csc_cdt}
	\end{equation}
	\end{widetext}
where $C_{i}$ is defined in (\ref{calCi_def}), and the non-vanishing Poisson brackets are given by
	\begin{equation}
	\left[p^{ij}\left(\vec{x}\right),\tilde{\pi}^{kl}\left(\vec{y}\right)\right]=\frac{1}{2N\left(\vec{y}\right)}\frac{\delta^{2}S_{B}}{\delta B_{ij}\left(\vec{x}\right)\delta B_{kl}\left(\vec{y}\right)},\label{SB_BB}
	\end{equation}
	\begin{equation}
	\left[\pi\left(\vec{x}\right),\tilde{\pi}^{kl}\left(\vec{y}\right)\right]=\frac{1}{2}\frac{\delta}{\delta N\left(\vec{x}\right)}\left(\frac{1}{N\left(\vec{y}\right)}\frac{\delta S_{B}}{\delta B_{kl}\left(\vec{y}\right)}\right),\label{=00005Bpi,pi^kl=00005D}
	\end{equation}
	\begin{eqnarray}
	 \left[\tilde{\pi}^{ij}\left(\vec{x}\right),\tilde{\pi}^{kl}\left(\vec{y}\right)\right] &= &\frac{1}{2N\left(\vec{y}\right)}\frac{\delta^{2}S_{B}}{\delta h_{ij}\left(\vec{x}\right)\delta B_{kl}\left(\vec{y}\right)}\nonumber \\
	 &  & -\frac{1}{2N\left(\vec{x}\right)}\frac{\delta^{2}S_{B}}{\delta B_{ij}\left(\vec{x}\right)\delta h_{kl}\left(\vec{y}\right)},
	\end{eqnarray}
	\begin{eqnarray}
	&  & \left[H_{\mathrm{C}},\tilde{\pi}^{ij}\left(\vec{x}\right)\right]\nonumber \\
	& = & -\frac{\delta S_{B}}{\delta h_{ij}(\vec{x})}+\int\mathrm{d}^{3}y\frac{N(\vec{y})}{N(\vec{x})}\frac{\delta^{2}S_{B}}{\delta B_{ij}(\vec{x})\delta h_{kl}(\vec{y})}B_{kl}(\vec{y}),\quad \label{T^ij}
	\end{eqnarray}
	\begin{eqnarray}
	\left[H_{\mathrm{C}},\pi(\vec{x})\right] & = & C'\left(\vec{x}\right)+2\tilde{\pi}^{ij}(\vec{x})B_{ij}(\vec{x})\approx C'\left(\vec{x}\right), \label{PB_HC_pi}
	\end{eqnarray}
where we define
	\begin{equation}
	C'\left(\vec{x}\right)\equiv-\frac{\delta S_{B}}{\delta N(\vec{x})}+\frac{1}{N(\vec{x})}\frac{\delta S_{B}}{\delta B_{ij}(\vec{x})}B_{ij}(\vec{x}),\label{C^prime}
	\end{equation}
In the following, we discuss the results of the consistency conditions (\ref{eq:csc_cdt}).

The first line in (\ref{eq:csc_cdt}) implies that
	\begin{equation}
	C_{i} \approx 0_{i},\label{C_i}
	\end{equation}
which are three secondary constraints. 
Thanks to the properties in (\ref{C_i property}), $C_{i}$ defined in (\ref{C_i}) must have vanishing Poisson brackets with any other constraints and thus be of the first class. 
For the second line in (\ref{eq:csc_cdt}), note  that (\ref{SB_BB}) is proportional to the kinetic matrix
	\begin{equation}
	\frac{\delta^{2}S_{B}}{\delta B_{ij}\left(\vec{x}\right)\delta B_{kl}\left(\vec{y}\right)},\label{S_BB}
	\end{equation}
which we assume to be non-degenerate. Mathematically, the non-degeneracy implies that (\ref{S_BB}) formally possesses an ``inverse'' $\mathcal{G}_{ij,kl}\left(\vec{x},\vec{y}\right)$
satisfying the relation
	\begin{equation}
	\int \mathrm{d}^{3}z\,\mathcal{G}_{ij,mn}\left(\vec{x},\vec{z}\right)\frac{\delta^{2}S_{B}}{\delta B_{mn}\left(\vec{z}\right)\delta B_{kl}\left(\vec{y}\right)}\equiv\mathbf{1}_{ij}^{kl}\delta^{3}\left(\vec{x}-\vec{y}\right),\label{inverse}
	\end{equation}
where $\mathcal{G}_{ij,kl}\left(\vec{x},\vec{y}\right)$ is symmetric in the sense that
	\begin{equation}
	\mathcal{G}_{ij,kl}\left(\vec{x},\vec{y}\right)=\mathcal{G}_{kl,ij}\left(\vec{y},\vec{x}\right),\label{inverse 2}
	\end{equation}
and $\mathbf{1}_{ij}^{kl}$ is the identity in the linear space of $3\times3$ symmetric matrices. 
Physically, if (\ref{S_BB}) is degenerate, the tensorial DoFs encoded in the spatial metric $h_{ij}$ would get lost, which is not the case we are concerned in the current paper. 
Therefore throughout this paper we assume that (\ref{S_BB}) is non-degenerate. 
Under this assumption, we can solve the Lagrange multiplier $\lambda_{kl}=0$.
This also leads to another secondary constraint
	\begin{equation}
	C' \approx 0,\label{eq:C^prime}
	\end{equation}
with $C'$ defined in (\ref{C^prime}), which arises in the third line in (\ref{eq:csc_cdt}).
For the last line of (\ref{eq:csc_cdt}), since (\ref{S_BB}) is not degenerate, we are able to  fix the Lagrange multiplier $\mu_{kl}$. Therefore there is no further secondary constraint.

The secondary constraints must be preserved in time evolution as well. 
Again, from (\ref{C_i property}) and (\ref{PB_Ci_HC}), the consistency condition for $C_{i}$ is automatically satisfied.
Before checking the consistency condition for  the secondary constraints $C' \approx 0$, we slightly rewrite the consistency conditions (\ref{eq:csc_cdt}) in order to simplify the calculations. 
This will also inspire ourselves on how to look for the TTDoF conditions for our theory.

In order to simplify the analysis, we concentrate on the sub-matrix in (\ref{eq:csc_cdt})
	\begin{equation}
	\begin{aligned}
	&  \mathcal{M}^{ab}\left(\vec{x},\vec{y}\right) \\
	 \equiv & \left(\begin{array}{ccc}
	0 & 0 & [p^{ij}(\vec{x}),\tilde{\pi}^{kl}(\vec{y})]\\
	0 & 0 & [\pi(\vec{x}),\tilde{\pi}^{kl}(\vec{y})]\\{}
	[\tilde{\pi}^{ij}(\vec{x}),p^{kl}(\vec{y})] & [\tilde{\pi}^{ij}(\vec{x}),\pi(\vec{y})] & [\tilde{\pi}^{ij}(\vec{x}),\tilde{\pi}^{kl}(\vec{y})]
	\end{array}\right). \label{M^ab}
	\end{aligned}
	\end{equation}
In the following we shall see that $\mathcal{M}^{ab}$ is degenerate in the sense that it possesses at least one null eigenvector $\mathcal{V}_{b}\left(\vec{y}\right)\neq0$ satisfying
	\begin{equation}
	\int \mathrm{d}^{3}y\,\mathcal{M}^{ab}\left(\vec{x},\vec{y}\right)\mathcal{V}_{b}\left(\vec{y}\right)\approx0^{a}.\label{MV}
	\end{equation}
To this end, we write the null eigenvector to be the form 
	\begin{equation}
	\mathcal{V}_{a}=\left(\begin{array}{c}
	X_{ij}\\
	Y\\
	Z_{ij}
	\end{array}\right), \label{null_vector}
	\end{equation}
and it immediately follows that $Z_{ij} = 0$, due to the same reason that the Lagrange multiplier $\lambda_{kl}=0$.
Thus (\ref{MV}) yields
	\begin{eqnarray}
	0 & \approx & \int\mathrm{d}^{3}y\Big(\left[\tilde{\pi}^{ij}\left(\vec{x}\right),p^{kl}\left(\vec{y}\right)\right]X_{kl}\left(\vec{y}\right)\nonumber \\
	&  & \qquad\qquad+\left[\tilde{\pi}^{ij}\left(\vec{x}\right),\pi\left(\vec{y}\right)\right]Y\left(\vec{y}\right)\Big).\label{MV_3}
	\end{eqnarray}
Plug (\ref{SB_BB}) and (\ref{=00005Bpi,pi^kl=00005D}) into (\ref{MV_3}), we have
	\begin{eqnarray}
	\int \mathrm{d}^{3}y\Bigg[\frac{1}{N\left(\vec{x}\right)}\frac{\delta^{2}S_{B}}{\delta B_{ij}\left(\vec{x}\right)\delta B_{kl}\left(\vec{y}\right)}X_{kl}\left(\vec{y}\right)\nonumber \\
	+\frac{\delta}{\delta N\left(\vec{y}\right)}\left(\frac{1}{N\left(\vec{x}\right)}\frac{\delta S_{B}}{\delta B_{ij}\left(\vec{x}\right)}\right)Y\left(\vec{y}\right)\Bigg] & \approx & 0,\label{MV_3-2}
	\end{eqnarray}
which yields the solution for $X_{kl}$ in terms of $Y$:
	\begin{equation}
	X_{kl}\left(\vec{y}\right)=\int \mathrm{d}^{3}z\,\chi_{kl}\left(\vec{y},\vec{z}\right)Y\left(\vec{z}\right),\label{X_kl}
	\end{equation}
where we define
	\begin{eqnarray}
	\chi_{kl}\left(\vec{y},\vec{z}\right) & \equiv & -\int\mathrm{d}^{3}x\,\mathcal{G}_{kl,ij}\left(\vec{y},\vec{x}\right)N\left(\vec{x}\right)\nonumber \\
	&  & \qquad\times\frac{\delta}{\delta N\left(\vec{z}\right)}\left(\frac{1}{N\left(\vec{x}\right)}\frac{\delta S_{B}}{\delta B_{ij}\left(\vec{x}\right)}\right), \label{chi}
	\end{eqnarray}
with $\mathcal{G}_{ij,kl}\left(\vec{x},\vec{y}\right)$ defined in (\ref{inverse}).
To conclude, the null eigenvector (\ref{null_vector}) can be written as
	\begin{equation}
	\mathcal{V}_{a}=\int \mathrm{d}^{3}y\left(\begin{array}{c}
	\chi_{ij}\left(\vec{x},\vec{y}\right)\\
	\delta^{3}\left(\vec{x}-\vec{y}\right)\\
	0_{ij}
	\end{array}\right)Y\left(\vec{y}\right),\label{null_vector-1}
	\end{equation}
where $\chi_{ij}$ is given in (\ref{chi}).

We can thus make a linear combination of $\pi$ and $p^{ij}$ by employing the null eigenvector $\mathcal{V}_{a}$ in (\ref{null_vector-1}),
	\begin{eqnarray}
	&  & \int\mathrm{d}^{3}z\,\varphi^{a}\left(\vec{z}\right)\mathcal{V}_{a}\left(\vec{z}\right)\nonumber \\
	& = & \int\mathrm{d}^{3}z\left(\begin{array}{ccc}
	p^{mn} & \pi & \tilde{\pi}^{mn}\end{array}\right)(\vec{z})\left(\begin{array}{c}
	X_{mn}\\
	Y\\
	0_{mn}
	\end{array}\right)(\vec{z})\nonumber \\
	& \equiv & \int\mathrm{d}^{3}z\,\tilde{\pi}\left(\vec{z}\right)Y\left(\vec{z}\right),\label{phiV}
	\end{eqnarray}
where we define
	\begin{equation}
	\tilde{\pi}\left(\vec{z}\right)\equiv\pi\left(\vec{z}\right)+\int \mathrm{d}^{3}y\,\chi_{mn}\left(\vec{z},\vec{y}\right)p^{mn}\left(\vec{z}\right)\approx0.\label{pi^tilde}
	\end{equation}

Now the set of independent primary constraints can be chosen to be $\{\pi_{i},\tilde{\pi}^{ij},\tilde{\pi},p^{ij}\}$. With this new set of primary constraints, the consistency conditions (\ref{eq:csc_cdt}) reduce to
	\begin{widetext}
	\begin{equation}
	\int \mathrm{d}^{3}y\left(\begin{array}{cccc}
	0 & 0 & 0 & 0\\
	0 & 0 & 0 & [p^{ij}(\vec{x}),\tilde{\pi}^{kl}(\vec{y})]\\
	0 & 0 & 0 & 0\\
	0 & [\tilde{\pi}^{ij}(\vec{x}),p^{kl}(\vec{y})] & 0 & [\tilde{\pi}^{ij}(\vec{x}),\tilde{\pi}^{kl}(\vec{y})]
	\end{array}\right)\left(\begin{array}{c}
	\lambda^{k}\\
	\mu_{kl}\\
	\lambda\\
	0_{kl}
	\end{array}\right)(\vec{y})\approx\left(\begin{array}{c}
	C_{i}(\vec{x})\\
	0^{ij}\\
	C^{\prime}(\vec{x})\\{}
	[H_{\mathrm{C}},\tilde{\pi}^{ij}(\vec{x})]
	\end{array}\right),\label{eq:csc_cdt-1}
	\end{equation}
	\end{widetext}
where we have used the fact that $[H_{\mathrm{C}},\tilde{\pi}] = [H_{\mathrm{C}},\pi] \equiv C'$.

In the above, we have shown that for a degenerate matrix of Poisson brackets, we may find its null eigenvector and use this null eigenvector to simplify the matrix of Poisson brackets. 
This is exactly the same trick have used in \cite{Gao:2018znj}, which we will also employ in order to find the TTDoF conditions in the following section.
The Point is that the matrix $\mathcal{M}^{ab}$ in (\ref{M^ab}) is always degenerate, while the matrix $\mathcal{W}^{ab}$ in (\ref{W^ab-2}) is not degenerate in general. As a result, some conditions have to be imposed in order to have the desired degeneracy, which are nothing but the TTDoF conditions we are looking for.

\section{The conditions for the two tensorial degrees of freedom}
\label{sec:TTcond}

\subsection{The null eigenvalue equations}

We are ready to check the consistency conditions for the secondary
constraints $C_{i} \approx 0$ and $C' \approx 0$. 
First of all, as has been mentioned in the above section, the consistency condition for $C_{i}$ is
	\begin{equation}
	\frac{\mathrm{d}{C}_{i}}{\mathrm{d}t}\left(\vec{x}\right)\equiv\left[C_{i}\left(\vec{x}\right),H_{\mathrm{T}}\right]\approx 0,
	\end{equation}
which is automatically satisfied according to the property in (\ref{C_i property}).
On the other hand, the consistency condition for $C'$ is 
	\begin{eqnarray}
	0\approx\frac{\mathrm{d}C^{\prime}}{\mathrm{d}t} & \equiv & \left[C^{\prime}\left(\vec{x}\right),H_{\mathrm{T}}\right]\nonumber \\
	& = & \left[C^{\prime}\left(\vec{x}\right),H_{\mathrm{C}}\right]+\int\mathrm{d}^{3}y\left[C^{\prime}\left(\vec{x}\right),p^{kl}\left(\vec{y}\right)\right]\mu_{kl}\left(\vec{y}\right)\nonumber \\
	&  & +\int\mathrm{d}^{3}y\left[C^{\prime}\left(\vec{x}\right),\tilde{\pi}\left(\vec{y}\right)\right]\lambda\left(\vec{y}\right).\label{cc_Cprime}
	\end{eqnarray}
If no further conditions are imposed, (\ref{cc_Cprime}) merely fixes the Lagrange multipliers and does not yield any further constraint. 

At this point, we may count the number of DoFs of our theory with. 
In total there are 20 constraints $\left\{ \tilde{\pi},\pi_{i},\tilde{\pi}^{ij},C^{\prime},C_{i},p^{ij}\right\} \approx 0$, of which the Poisson brackets can be summarized in the so-called Dirac matrix:
	\begin{widetext}
	\begin{center}
		\begin{tabular}{c|cccccc}
			\hline 
			$[\cdot,\cdot]$ & $\pi_{k}(\vec{y})$ & $C_{k}(\vec{y})$ & $p^{kl}(\vec{y})$ & $\tilde{\pi}(\vec{y})$ & $\tilde{\pi}^{kl}(\vec{y})$ & $C'(\vec{y})$\tabularnewline
			\hline 
			$\pi_{i}(\vec{x})$ & 0 & 0 & 0 & 0 & 0 & 0\tabularnewline
			$C_{i}(\vec{x})$ & 0 & 0 & 0 & 0 & 0 & 0\tabularnewline
			$p^{ij}(\vec{x})$ & 0 & 0 & 0 & 0 & $[p^{ij}(\vec{x}),\tilde{\pi}^{kl}(\vec{y})]$ & $[p^{ij}(\vec{x}),C'(\vec{y})]$\tabularnewline
			$\tilde{\pi}(\vec{x})$ & 0 & 0 & 0 & 0 & 0 & $[\tilde{\pi}(\vec{x}),C'(\vec{y})]$\tabularnewline
			$\tilde{\pi}^{ij}(\vec{x})$ & 0 & 0 & $[\tilde{\pi}^{ij}(\vec{x}),p^{kl}(\vec{y})]$ & 0 & $[\tilde{\pi}^{ij}(\vec{x}),\tilde{\pi}^{kl}(\vec{y})]$ & $[\tilde{\pi}^{ij}(\vec{x}),C'(\vec{y})]$\tabularnewline
			$C'(\vec{x})$ & 0 & 0 & $[C'(\vec{x}),p^{kl}(\vec{y})]$ & $[C'(\vec{x}),\tilde{\pi}(\vec{y})]$ & $[C'(\vec{x}),\tilde{\pi}^{kl}(\vec{y})]$ & 0\tabularnewline
			\hline 
		\end{tabular}
		\par\end{center}
	\end{widetext}
From the above matrix, $\pi_{i}\approx0$ and $C_{i}\approx0$ are first class constraints by the terminology of Dirac \cite{Henneaux:1992ig}, which correspond to the spatial diffeomorphism of the theory. 
All the other constraints are second class. 
The number of DoFs is thus
	\begin{eqnarray}
	\#_{\text{DoF}} & = & \frac{1}{2}\left(\#_{\text{var}}\times 2-\#_{\text{1st}}\times 2-\#_{\text{2nd}}\right)\nonumber \\
	 & = & \frac{1}{2}\left(16\times 2-6\times 2-14\right)\nonumber \\
	 & = & 3.
	\end{eqnarray}
All of the above results are consistent with the previous studies \cite{Gao:2014fra,Saitou:2016lvb}, as expected.

For our purpose, we need extra first or second class constraints to reduce the DoFs from three to two. 
This requirement translates to requiring that the sub-matrix (by omitting the first two columns and rows in the above Dirac matrix)
	\begin{widetext}
	\begin{equation}
	\mathcal{W}^{ab}\left(\vec{x},\vec{y}\right)\equiv\left(\begin{array}{cccc}
	0 & 0 & [p^{ij}(\vec{x}),\tilde{\pi}^{kl}(\vec{y})] & [p^{ij}(\vec{x}),C^{\prime}(\vec{y})]\\
	0 & 0 & 0 & [\tilde{\pi}(\vec{x}),C^{\prime}(\vec{y})]\\{}
	[\tilde{\pi}^{ij}(\vec{x}),p^{kl}(\vec{y})] & 0 & [\tilde{\pi}^{ij}(\vec{x}),\tilde{\pi}^{kl}(\vec{y})] & [\tilde{\pi}^{ij}(\vec{x}),C^{\prime}(\vec{y})]\\{}
	[C^{\prime}(\vec{x}),p^{kl}(\vec{y})] & [C^{\prime}(\vec{x}),\tilde{\pi}(\vec{y})] & [C^{\prime}(\vec{x}),\tilde{\pi}^{kl}(\vec{y})] & 0
	\end{array}\right), \label{W^ab-2}
	\end{equation}
	\end{widetext}
has to be degenerated. 
As what we have discussed in Sec. \ref{subsec:The-consistency-condition}, we  assume that there exists a non-trivial null eigenvector for (\ref{W^ab-2})
	\begin{equation}
	\mathcal{U}_{b}\equiv\left(\begin{array}{c}
	W_{kl}\\
	W\\
	U_{kl}\\
	U
	\end{array}\right)\neq0_{b},\label{U_b}
	\end{equation}
such that
	\begin{equation}
	\int \mathrm{d}^{3}y\,\mathcal{W}^{ab}\left(\vec{x},\vec{y}\right)\mathcal{U}_{b}\left(\vec{y}\right)\approx0^{a}.\label{null equation}
	\end{equation}
More precisely, (\ref{null equation}) can be split into
	\begin{eqnarray}
	&  & \int\mathrm{d}^{3}y\Big([p^{ij}(\vec{x}),\tilde{\pi}^{kl}(\vec{y})]U_{kl}\left(\vec{y}\right)\nonumber \\
	&  & \qquad\quad+[p^{ij}(\vec{x}),C'(\vec{y})]U\left(\vec{y}\right)\Big)\approx 0^{ij},\label{WU_1-2}
	\end{eqnarray}
	\begin{equation}
	\int \mathrm{d}^{3}y[\tilde{\pi}(\vec{x}),C^{\prime}(\vec{y})]U\left(\vec{y}\right)\approx 0,\label{WU_2-2}
	\end{equation}
	\begin{eqnarray}
	 &  & \int \mathrm{d}^{3}y\,\Big([\tilde{\pi}^{ij}(\vec{x}),p^{kl}(\vec{y})]W_{kl}\left(\vec{y}\right)+[\tilde{\pi}^{ij}(\vec{x}),\tilde{\pi}^{kl}(\vec{y})]U_{kl}\left(\vec{y}\right)\nonumber \\
	 &  &  \qquad \quad+[\tilde{\pi}^{ij}(\vec{x}),C^{\prime}(\vec{y})]U\left(\vec{y}\right)\Big)\approx 0^{ij},\label{WU_3-2}
	\end{eqnarray}
	\begin{eqnarray}
	 &  & \int \mathrm{d}^{3}y\Big([C^{\prime}(\vec{x}),p^{kl}(\vec{y})]W_{kl}\left(\vec{y}\right)+[C^{\prime}(\vec{x}),\tilde{\pi}(\vec{y})]W\left(\vec{y}\right)\nonumber \\
	 &  & \qquad \quad +[C^{\prime}(\vec{x}),\tilde{\pi}^{kl}(\vec{y})]U_{kl}\left(\vec{y}\right)\Big)\approx0.\label{WU_4}
	\end{eqnarray}
By formally solving the null eigenvector from the above null eigenvalue equation, we can find the conditions to remove the scalar DoF, which we shall shown in the next two sections.

\subsection{The first TTDoF condition}

According to whether $U$ in (\ref{U_b}) identically vanishes or not, there are two cases:
\begin{itemize}
	\item If $U\neq0$, it immediately follows from (\ref{WU_2-2}) that we need to require
	\begin{equation}
	\left[\tilde{\pi}\left(\vec{x}\right),C^{\prime}\left(\vec{y}\right)\right]\approx 0.\label{PB_pitld_Cprime}
	\end{equation}
	\item If $U\equiv0$, from (\ref{WU_1-2}) we get $U_{kl} = 0$, then (\ref{WU_3-2}) also implies that $W_{kl} = 0$. Thus according to (\ref{WU_4}), in order to have a non-trivial solution for $\mathcal{U}_{b}$, we have to require
	\begin{equation}
	\left[C^{\prime}\left(\vec{x}\right),\tilde{\pi}\left(\vec{y}\right)\right]\approx 0,\label{PB_Cprime_pitld}
	\end{equation}
	such that the non-trivial solution takes the form
	\begin{equation}
	\mathcal{U}_{b}^{\left(1\right)}\equiv\left(\begin{array}{c}
	0_{kl}\\
	W\\
	0_{kl}\\
	0
	\end{array}\right),\label{U_b-1}
	\end{equation}
	with $W\neq 0$.
\end{itemize}
It is interesting to note that in both cases, (\ref{PB_pitld_Cprime}) (or equivalently (\ref{PB_Cprime_pitld})) arises as a necessary condition for the degeneracy of $\mathcal{W}^{ab}(\vec{x},\vec{y})$ defined in (\ref{W^ab-2}).

By evaluating the Poisson bracket explicitly,  (\ref{PB_pitld_Cprime}) is equivalent to requiring
	\begin{equation}
	\mathcal{S}\left(\vec{x},\vec{y}\right)\approx0,\label{1st-cdt}
	\end{equation}
where
	\begin{widetext}
	\begin{eqnarray}
	\mathcal{S}\left(\vec{x},\vec{y}\right) & \equiv & \frac{\delta^{2}S_{B}}{\delta N\left(\vec{x}\right)\delta N\left(\vec{y}\right)}-\int \mathrm{d}^{3}x^{\prime}\int \mathrm{d}^{3}y^{\prime}N\left(\vec{x}^{\prime}\right)\frac{\delta}{\delta N\left(\vec{x}\right)}\left(\frac{1}{N\left(\vec{x}^{\prime}\right)}\frac{\delta S_{B}}{\delta B_{i^{\prime}j^{\prime}}\left(\vec{x}^{\prime}\right)}\right)\nonumber \\
	 &  & \times\mathcal{G}_{i^{\prime}j^{\prime},k^{\prime}l^{\prime}}\left(\vec{x}^{\prime},\vec{y}^{\prime}\right)N\left(\vec{y}^{\prime}\right)\frac{\delta}{\delta N\left(\vec{y}\right)}\left(\frac{1}{N\left(\vec{y}^{\prime}\right)}\frac{\delta S_{B}}{\delta B_{k^{\prime}l^{\prime}}\left(\vec{y}^{\prime}\right)}\right).\label{S_xpl}
	\end{eqnarray}
	\end{widetext}
From (\ref{S_xpl}), $\mathcal{S}$ is symmetric in the sense that $\mathcal{S}(\vec{x},\vec{y}) = \mathcal{S}(\vec{y},\vec{x})$.
See Appendix \ref{sec:The-first-TTDoF} for more details in deriving (\ref{S_xpl}).
The action $S_{B}$ (\ref{S_B}) must satisfy  the equation (\ref{1st-cdt}) in order to have $\mathcal{W}^{ab}$ in  (\ref{W^ab-2}) being degenerate.
We dub (\ref{1st-cdt}) as the first TTDoF condition in our formalism.

From the expression in (\ref{S_xpl}), the physical meaning of the first TTDoF condition (\ref{1st-cdt}) is transparent, which implies that the $\left\{ N,K_{ij}\right\}$-sector has to be degenerate, if we want to eliminate the scalar mode.

At this point, we have shown that as long as the first TTDoF condition is satisfied, $\mathcal{W}^{ab}$ is degenerate and there is a null eigenvector $\mathcal{U}_{b}^{\left(1\right)}$, which is in the form (\ref{U_b-1}).
However, if $\mathcal{U}_{b}^{\left(1\right)}$ is the only null eigenvector for $\mathcal{W}^{ab}$, we will be left with an odd number of second class constraints. As a result, the the dimensions of the phase space at each spacetime point will be odd, which may lead to inconsistency \cite{Henneaux:2009zb,Li:2009bg,Gao:2018znj}.
In other words, the scalar DoF is not eliminated completely.
To avoid this situation, we have to require that $\mathcal{W}^{ab}$ possesses a second null vector $\mathcal{U}_{b}^{\left(2\right)}$, which we shall discuss below.

\subsection{The second TTDoF condition \label{subsec:The-second-TTDoF}}

In order to have a second null eigenvector, we have to assume $U\neq 0 $.
In this case,  (\ref{WU_1-2})-(\ref{WU_4}) are reduced to be
	\begin{eqnarray}
	&  & \int \mathrm{d}^{3}y\,\Big([p^{ij}(\vec{x}),\tilde{\pi}^{kl}(\vec{y})]U_{kl}\left(\vec{y}\right)\nonumber \\
	&  & \qquad\quad+[p^{ij}(\vec{x}),C^{\prime}(\vec{y})]U\left(\vec{y}\right)\Big)\approx0^{ij},\label{WU_1}
	\end{eqnarray}
	\begin{eqnarray}
	 &  & \int \mathrm{d}^{3}y\,\Bigg([\tilde{\pi}^{ij}(\vec{x}),p^{kl}(\vec{y})]W_{kl}\left(\vec{y}\right)+[\tilde{\pi}^{ij}(\vec{x}),\tilde{\pi}^{kl}(\vec{y})]U_{kl}\left(\vec{y}\right)\nonumber \\
	 &  & \qquad \quad +[\tilde{\pi}^{ij}(\vec{x}),C^{\prime}(\vec{y})]U\left(\vec{y}\right)\Bigg)\approx0^{ij},\label{WU_2}
	\end{eqnarray}
and
	\begin{eqnarray}
	&  & \int\mathrm{d}^{3}y\,\Big([C^{\prime}(\vec{x}),p^{kl}(\vec{y})]W_{kl}\left(\vec{y}\right)\nonumber \\
	&  & \qquad\quad+[C^{\prime}(\vec{x}),\tilde{\pi}^{kl}(\vec{y})]U_{kl}\left(\vec{y}\right)\Big)\approx0.\label{WU_3}
	\end{eqnarray}
Following the similar steps in Sec. \ref{subsec:The-consistency-condition}, after tedious but straightforward calculations, we can solve the second null eigenvector to be
	\begin{equation}
	\mathcal{U}_{b}^{\left(2\right)}\left(\vec{z}\right)\equiv\left(\begin{array}{c}
	W_{mn}\\
	0\\
	U_{mn}\\
	U
	\end{array}\right)\left(\vec{z}\right)=\int \mathrm{d}^{3}y\left(\begin{array}{c}
	w_{mn}\left(\vec{z},\vec{y}\right)\\
	0\\
	u_{mn}\left(\vec{z},\vec{y}\right)\\
	\delta^{3}\left(\vec{z}-\vec{y}\right)
	\end{array}\right)U\left(\vec{y}\right),\label{2nd_null_vct}
	\end{equation}
where we have fixed $W=0$ such that $\mathcal{U}_{b}^{(2)}$ and $\mathcal{U}_{b}^{(1)}$ defined in (\ref{U_b-1}) are orthogonal to each other.
 In (\ref{2nd_null_vct}), we define
	\begin{equation}
	u_{mn}\left(\vec{z},\vec{y}\right)\equiv\int \mathrm{d}^{3}xN\left(\vec{x}\right)\mathcal{G}_{mn,ij}\left(\vec{z},\vec{x}\right)\frac{\delta C^{\prime}\left(\vec{y}\right)}{\delta B_{ij}\left(\vec{x}\right)},\label{u_mn}
	\end{equation}
and
	\begin{eqnarray}
	&  & w_{mn}\left(\vec{z},\vec{y}\right)\equiv-\int\mathrm{d}^{3}x^{\prime}N\left(\vec{x}^{\prime}\right)\mathcal{G}_{mn,i^{\prime}j^{\prime}}\left(\vec{z},\vec{x}^{\prime}\right)\nonumber \\
	&  & \qquad \times\Bigg[\frac{\delta C^{\prime}\left(\vec{y}\right)}{\delta h_{i^{\prime}j^{\prime}}\left(\vec{x}^{\prime}\right)}+\int\mathrm{d}^{3}y^{\prime}\Bigg(\frac{1}{N(\vec{x}^{\prime})}\frac{\delta^{2}S_{B}}{\delta h_{k^{\prime}l^{\prime}}(\vec{y}^{\prime})\delta B_{i^{\prime}j^{\prime}}(\vec{x}^{\prime})}\nonumber \\
	&  & \qquad\quad -\frac{1}{N(\vec{y}^{\prime})}\frac{\delta^{2}S_{B}}{\delta h_{i^{\prime}j^{\prime}}(\vec{x}^{\prime})\delta B_{k^{\prime}l^{\prime}}(\vec{y}^{\prime})}\Bigg)u_{k^{\prime}l^{\prime}}\left(\vec{y}^{\prime},\vec{y}\right)\Bigg].\label{w_mn}
	\end{eqnarray}
Substituting (\ref{u_mn}) and (\ref{w_mn}) into (\ref{WU_3}), we
find the second TTDoF condition
	\begin{equation}
	\mathcal{J}\left(\vec{x},\vec{y}\right)\approx0,\label{2nd_cdt}
	\end{equation}
where
	\begin{widetext}
	\begin{eqnarray}
	\mathcal{J}\left(\vec{x},\vec{y}\right) & \equiv & \int\mathrm{d}^{3}x^{\prime}\int\mathrm{d}^{3}y^{\prime}\int\mathrm{d}^{3}x^{\prime\prime}\int\mathrm{d}^{3}y^{\prime\prime}\frac{\delta C^{\prime}\left(\vec{x}\right)}{\delta B_{ij}\left(\vec{x}^{\prime}\right)}\mathcal{G}_{ij,i^{\prime}j^{\prime}}\left(\vec{x}^{\prime},\vec{x}^{\prime\prime}\right)\nonumber \\
	&  & \quad\times N\left(\vec{x}^{\prime\prime}\right)\frac{\delta^{2}S_{B}}{\delta h_{i^{\prime}j^{\prime}}\left(\vec{x}^{\prime\prime}\right)\delta B_{k^{\prime}l^{\prime}}\left(\vec{y}^{\prime\prime}\right)}\mathcal{G}_{k^{\prime}l^{\prime},kl}\left(\vec{y}^{\prime\prime},\vec{y}^{\prime}\right)\frac{\delta C^{\prime}\left(\vec{y}\right)}{\delta B_{kl}\left(\vec{y}^{\prime}\right)}\nonumber \\
	&  & -\int\mathrm{d}^{3}x^{\prime}\int\mathrm{d}^{3}y^{\prime}\frac{\delta C^{\prime}\left(\vec{x}\right)}{\delta B_{ij}\left(\vec{x}^{\prime}\right)}\mathcal{G}_{ij,kl}\left(\vec{x}^{\prime},\vec{y}^{\prime}\right)N\left(\vec{y}^{\prime}\right)\frac{\delta C^{\prime}\left(\vec{y}\right)}{\delta h_{kl}\left(\vec{y}^{\prime}\right)}-\left(\vec{x}\leftrightarrow\vec{y}\right). \label{J_xpl}
	\end{eqnarray}
	\end{widetext}
Note $\mathcal{J}(\vec{x},\vec{y})$ is antisymmetric in the sense that $\mathcal{J}(\vec{y},\vec{x}) = - \mathcal{J}(\vec{x},\vec{y})$.

As we shall see below, if both the conditions (\ref{1st-cdt}) and (\ref{2nd_cdt}) are satisfied, there will be two degrees of freedom propagating in our theory.

\subsection{Counting the degrees of freedom}

We can make linear combinations of the constraints by employing the null eigenvectors (similar to (\ref{phiV})).
For the first null eigenvector $\mathcal{U}_{b}^{(1)}$ in (\ref{U_b-1}), we simply get
	\begin{eqnarray}
	&  & \int\mathrm{d}^{3}y\left(\begin{array}{cccc}
	p^{mn} & \tilde{\pi} & \tilde{\pi}^{mn} & C^{\prime}\end{array}\right)\left(\begin{array}{c}
	0_{mn}\\
	W\\
	0_{mn}\\
	0
	\end{array}\right)\nonumber \\
	& \equiv & \int\mathrm{d}^{3}y\,\tilde{\pi}(\vec{y})\,W(\vec{y}) \approx 0,
	\end{eqnarray}
which trivially implies $\tilde{\pi} \approx 0$ and is nothing new.
For the second null eigenvector $\mathcal{U}_{b}^{(2)}$ given in (\ref{2nd_null_vct}), we get
	\begin{eqnarray}
	&  & \int\mathrm{d}^{3}y\left(\begin{array}{cccc}
	p^{mn} & \tilde{\pi} & \tilde{\pi}^{mn} & C^{\prime}\end{array}\right)\left(\begin{array}{c}
	W_{mn}\\
	0\\
	U_{mn}\\
	U
	\end{array}\right)\nonumber \\
	& \equiv & \int\mathrm{d}^{3}y\,\tilde{C}(\vec{y})\,U(\vec{y}).\label{Ctld}
	\end{eqnarray}
where we define
	\begin{eqnarray}
	\tilde{C}\left(\vec{y}\right) & \equiv & C'\left(\vec{y}\right)+\int\mathrm{d}^{3}z\,w_{mn}\left(\vec{z},\vec{y}\right)p^{mn}\left(\vec{z}\right)\nonumber \\
	&  & +\int\mathrm{d}^{3}z\,u_{mn}\left(\vec{z},\vec{y}\right)\tilde{\pi}^{mn}\left(\vec{z}\right),
	\end{eqnarray}
which is a linear combination of $C'$, $p^{ij}$ and $\tilde{\pi}^{ij}$.
Under the two TTDoF conditions (\ref{1st-cdt}) and (\ref{2nd_cdt}), the consistency condition for the new constraint $\tilde{C}\approx0$ yields 
	\begin{equation}
	\frac{\mathrm{d}\tilde{C}}{\mathrm{d}t}=\left[\tilde{C},H_{\mathrm{T}}\right]\approx\left[\tilde{C},H_{\mathrm{C}}\right]\approx 0,
	\end{equation}
which may or may not be satisfied automatically. In the later case, there will arise a tertiary constraint.
In the following, we count the number of DoFs in two cases according to whether the consistency condition for $\tilde{C}$ yields a tertiary constraint or not.

In the special case, if 
	\begin{equation}
	\left[\tilde{C}\left(\vec{x}\right),H_{\mathrm{C}}\right]\approx0\label{tertiary_cst}
	\end{equation}
is automatically satisfied in the subspace $\Gamma_{P}$, i.e., there is no tertiary constraint,  the Dirac matrix is 
	\begin{center}
		\begin{tabular}{c|cccccc}
			\hline 
			$\left[\cdot,\cdot\right]$ & $\pi_{k}$ & $C_{k}$ & $p^{kl}$ & $\tilde{\pi}$ & $\tilde{\pi}^{kl}$ & $\tilde{C}$\tabularnewline
			\hline 
			$\pi_{i}$ & 0 & 0 & 0 & 0 & 0 & 0\tabularnewline
			$C_{i}$ & 0 & 0 & 0 & 0 & 0 & 0\tabularnewline
			$p^{ij}$ & 0 & 0 & 0 & 0 & $[p^{ij}(\vec{x}),\tilde{\pi}^{kl}(\vec{y})]$ & 0\tabularnewline
			$\tilde{\pi}$ & 0 & 0 & 0 & 0 & 0 & 0\tabularnewline
			$\tilde{\pi}^{ij}$ & 0 & 0 & $[\tilde{\pi}^{ij}(\vec{x}),p^{kl}(\vec{y})]$ & 0 & $[\tilde{\pi}^{ij}(\vec{x}),\tilde{\pi}^{kl}(\vec{y})]$ & 0\tabularnewline
			$\tilde{C}$ & 0 & 0 & 0 & 0 & 0 & 0\tabularnewline
			\hline 
		\end{tabular}
		\par\end{center}
which means $\left\{ \pi_{i},C_{i},\tilde{\pi},\tilde{C}\right\} \approx0$
are 8 first class constraints and $\left\{ p^{ij},\tilde{\pi}^{ij}\right\} \approx 0$
are 12 second class constraints. In this case, the number of DoFs is simply given by
	\begin{eqnarray}
	\#_{\mathrm{dof}} & = & \frac{1}{2}\left(\#_{\mathrm{var}}\times2-\#_{\mathrm{1st}}\times2-\#_{\mathrm{2nd}}\right)\nonumber \\
	 & = & \frac{1}{2}\left(16\times2-8\times2-12\right)\nonumber \\
	 & = & 2.
	\end{eqnarray}
	
Generally,
	\begin{equation}
	\Phi\equiv\left[\tilde{C}\left(\vec{x}\right),H_{\mathrm{C}}\right]
	\end{equation}
does not vanish weakly, and thus $\Phi \approx 0$ is a tertiary constraint. In this case, the Dirac matrix is 
	\begin{widetext}
	\begin{center}
		\begin{tabular}{c|ccccccc}
			\hline 
			$\left[\cdot,\cdot\right]$ & $\pi_{k}$ & $C_{k}$ & $p^{kl}$ & $\tilde{\pi}$ & $\tilde{\pi}^{kl}$ & $\tilde{C}$ & $\Phi$\tabularnewline
			\hline 
			$\pi_{i}$ & 0 & 0 & 0 & 0 & 0 & 0 & 0\tabularnewline
			$C_{i}$ & 0 & 0 & 0 & 0 & 0 & 0 & 0\tabularnewline
			$p^{ij}$ & 0 & 0 & 0 & 0 & $[p^{ij}(\vec{x}),\tilde{\pi}^{kl}(\vec{y})]$ & 0 & $\ast$\tabularnewline
			$\tilde{\pi}$ & 0 & 0 & 0 & 0 & 0 & 0 & $[\tilde{\pi}(\vec{x}),\Phi(\vec{y})]$\tabularnewline
			$\tilde{\pi}^{ij}$ & 0 & 0 & $[\tilde{\pi}^{ij}(\vec{x}),p^{kl}(\vec{y})]$ & 0 & $[\tilde{\pi}^{ij}(\vec{x}),\tilde{\pi}^{kl}(\vec{y})]$ & 0 & $\ast$\tabularnewline
			$\tilde{C}$ & 0 & 0 & 0 & 0 & 0 & 0 & $[\tilde{C}(\vec{x}),\Phi(\vec{y})]$\tabularnewline
			$\Phi$ & 0 & 0 & $\ast$ & $[\Phi(\vec{x}),\tilde{\pi}(\vec{y})]$ & $\ast$ & $[\Phi(\vec{x}),\tilde{C}(\vec{y})]$ & $\ast$\tabularnewline
			\hline 
		\end{tabular}
		\par\end{center}
	\end{widetext}
In the above, entries with the symbol ``$\ast$'' represent Poisson brackets which are irrelevant to counting the number of DoFs. 
Apparently, there are 6 first class constraints together with 15 second-class constraints.
However, it is easy to show that the above Dirac matrix possesses 7 (instead of 6) null eigenvectors, if the two Possion brackets
	\begin{equation}
	[\tilde{\pi}(\vec{x}),\Phi(\vec{y})] ,\qquad[\tilde{C}(\vec{x}),\Phi(\vec{y})] \label{PB_Phi_two}
	\end{equation}
do not vanish simultaneously.
In this case, there are 7 first class constraints and 14 second class constraints. As a result, the number of DoFs is counted as
	\begin{eqnarray}
	\#_{\mathrm{dof}} & = & \frac{1}{2}\left(\#_{\mathrm{var}}\times2-\#_{\mathrm{1st}}\times2-\#_{\mathrm{2nd}}\right)\nonumber \\
	 & = & \frac{1}{2}\left(16\times2-7\times2-14\right)\nonumber \\
	 & = & 2.
	\end{eqnarray}
On the other hand, if both Poisson brackets in (\ref{PB_Phi_two}) vanish weakly, there will be more constraints and the number of DoFs will be less than two, which is not the case we are concerned.

To summarize, as long as the two TTDoF conditions (\ref{1st-cdt}) and (\ref{2nd_cdt}) are satisfied simultaneously for our theory (\ref{S_SCG I}),  the scalar DoF is suppressed completely, and we are left with only two tensorial DoFs.

\section{Applications} \label{sec:app}

For a general action (\ref{S_SCG I}), solving the two TTDoF conditions (\ref{1st-cdt}) and (\ref{2nd_cdt}) could be complicated, if not impossible.
In this section, we consider some special cases, and compare our results with the previous works on Lorentz breaking scalar-tensor theories with two DoFs.
These include the ``cuscuton theory'' \cite{Afshordi:2006ad} and its extensions \cite{Iyonaga:2018vnu} as well as the ``minimally modified gravity'' proposed in \cite{Lin:2017oow}.

\subsection{No spatial derivatives} \label{subsec:The-extended-cuscuton}

One of the complexities in solving the first and the second TTDoF conditions is the presence of spatial derivatives in the Lagrangian. A simpler situation is that the action does not involve any spatial derivatives, which is thus
	\begin{equation}
	S^{\left(\mathrm{n.d.}\right)}=\int \mathrm{d}t\mathrm{d}^{3}x\,N\sqrt{h}\,\mathcal{L}^{\left(\mathrm{n.d.}\right)}\left(N,h_{ij},K_{ij},R_{ij};t\right).\label{S_noder}
	\end{equation}
When considering the generally covariant scalar-tensor theories involving up to the second derivatives of the scalar field, the action in the unitary gauge with $\phi = t$ exactly falls into the form of (\ref{S_noder}).  

For the action (\ref{S_noder}), the first TTDoF condition reduces to be
	\begin{equation}
	\frac{\partial^{2}\left(N\mathcal{L}_{B}^{\left(\mathrm{n.d.}\right)}\right)}{\partial N^{2}}-\frac{\partial^{2}\mathcal{L}_{B}^{\left(\mathrm{n.d.}\right)}}{\partial N\partial B_{ij}}N\mathcal{G}_{ij,kl}\frac{\partial^{2}\mathcal{L}_{B}^{\left(\mathrm{n.d.}\right)}}{\partial N\partial B_{kl}}\approx 0,\label{ec_1st_cdt}
	\end{equation}
where $\mathcal{L}_{B}^{\left(\mathrm{n.d.}\right)}$ is $\mathcal{L}^{\left(\mathrm{n.d.}\right)}$ with $K_{ij}$ replaced by $B_{ij}$.
One class of second order scalar-tensor theory with two tensorial degrees of freedom is the cuscuton theory \cite{Afshordi:2006ad}. An extended cuscuton theory was studied in \cite{Iyonaga:2018vnu}, where a necessary condition for removing the scalar DoF was derived.
It is interesting to note that the necessary condition derived in \cite{Iyonaga:2018vnu} is nothing but the first TTDoF condition (\ref{ec_1st_cdt}), which by itself is a special case of the general form (\ref{1st-cdt}) and (\ref{S_xpl}) derived in this work. 

On the other hand, one of the main findings in this work is that the first TTDoF condition (\ref{ec_1st_cdt}) is not sufficient to remove the whole scalar DoF. 
The second TTDoF (\ref{2nd_cdt}) is also required in order to fully get rid of the scalar DoF. 
For the action (\ref{S_noder}), the second TTDoF condition (\ref{J_xpl}) is simplified to be 
	\begin{eqnarray}
	 &  & \int \mathrm{d}^{3}x\sqrt{h}\left(\alpha\nabla_{p}\beta-\beta\nabla_{p}\alpha\right)\nonumber \\
	 &  & \qquad \times\left(\Pi_{mn}^{pq}\nabla_{q}\Xi_{mn}-\Xi_{mn}\nabla_{q}\Pi_{mn}^{pq}\right)\approx 0,\label{ec_2nd_cdt_test_fct}
	\end{eqnarray}
where
	\begin{equation}
	\Xi_{mn}\equiv N\frac{\partial^{2}\mathcal{L}_{B}^{\left(\mathrm{E.C.}\right)}}{\partial N\partial B_{k^{\prime}l^{\prime}}}\mathcal{G}_{k^{\prime}l^{\prime},kl}\frac{\partial^{2}\mathcal{L}_{B}^{\left(\mathrm{E.C.}\right)}}{\partial B_{kl}\partial R_{mn}}-\frac{\partial^{2}\left(N\mathcal{L}_{B}^{\left(\mathrm{E.C.}\right)}\right)}{\partial N\partial R_{mn}},\label{Xi}
	\end{equation}
	\begin{equation}
	\Pi_{mn}^{pq}\equiv\mathcal{A}_{mn}^{ij,pq}\left(N\frac{\partial^{2}\mathcal{L}_{B}^{\left(\mathrm{E.C.}\right)}}{\partial N\partial B_{i^{\prime}j^{\prime}}}\mathcal{G}_{i^{\prime}j^{\prime},ij}-B_{ij}\right),\label{Pi}
	\end{equation}
	\begin{equation}
	\mathcal{A}_{kl}^{ij,mn}=h^{mp}\Sigma_{pkl}^{nij}-h^{pq}\Sigma_{pq(k}^{nij}h_{l)}^{m},\label{A}
	\end{equation}
and
	\begin{equation}
	\Sigma_{nkl}^{mij}=\frac{1}{2}\left(h_{k}^{m}h_{l}^{(i}h_{n}^{j)}+h_{l}^{m}h_{n}^{(i}h_{k}^{j)}-h_{n}^{m}h_{k}^{(i}h_{l}^{j)}\right).
	\end{equation}
In (\ref{ec_2nd_cdt_test_fct}), We have introduced the test functions $\alpha\left(\vec{x}\right)$ and $\beta\left(\vec{x}\right)$ in order to avoid dealing with the derivatives of the delta function.

A sufficient condition satisfying (\ref{ec_2nd_cdt_test_fct}) is
	\begin{equation}
	\Pi_{mn}^{pq}\nabla_{q}\Xi_{mn}-\Xi_{mn}\nabla_{q}\Pi_{mn}^{pq}\approx 0.\label{sffc_2nd_cdt_ec}
	\end{equation}
By plugging (\ref{Xi}) and (\ref{Pi}) into
(\ref{sffc_2nd_cdt_ec}), finally we get
	\begin{widetext}
	\begin{eqnarray}
	0 & \approx & \mathcal{A}_{mn}^{ij,pq}\left(N\frac{\partial^{2}\mathcal{L}_{B}^{\left(\mathrm{E.C.}\right)}}{\partial N\partial B_{i^{\prime}j^{\prime}}}\mathcal{G}_{i^{\prime}j^{\prime},ij}-B_{ij}\right)\nabla_{q}\left(N\frac{\partial^{2}\mathcal{L}_{B}^{\left(\mathrm{E.C.}\right)}}{\partial N\partial B_{k^{\prime}l^{\prime}}}\mathcal{G}_{k^{\prime}l^{\prime},kl}\frac{\partial^{2}\mathcal{L}_{B}^{\left(\mathrm{E.C.}\right)}}{\partial B_{kl}\partial R_{mn}}-\frac{\partial^{2}N\mathcal{L}_{B}^{\left(\mathrm{E.C.}\right)}}{\partial N\partial R_{mn}}\right)\nonumber \\
	 &  & -\left(N\frac{\partial^{2}\mathcal{L}_{B}^{\left(\mathrm{E.C.}\right)}}{\partial N\partial B_{k^{\prime}l^{\prime}}}\mathcal{G}_{k^{\prime}l^{\prime},kl}\frac{\partial^{2}\mathcal{L}_{B}^{\left(\mathrm{E.C.}\right)}}{\partial B_{kl}\partial R_{mn}}-\frac{\partial^{2}N\mathcal{L}_{B}^{\left(\mathrm{E.C.}\right)}}{\partial N\partial R_{mn}}\right)\nabla_{q}\mathcal{A}_{mn}^{ij,pq}\left(N\frac{\partial^{2}\mathcal{L}_{B}^{\left(\mathrm{E.C.}\right)}}{\partial N\partial B_{i^{\prime}j^{\prime}}}\mathcal{G}_{i^{\prime}j^{\prime},ij}-B_{ij}\right).\label{ec_2nd_cdt}
	\end{eqnarray}
	\end{widetext}
We will use (\ref{ec_2nd_cdt}) to find a concrete action as its application in Sec. \ref{subsec:The-quadratic-case}.

\subsection{Linear in the lapse}

Another interesting simplification is to assume that the action is linear in the lapse function, which is
	\begin{equation}
	S^{\left(\mathrm{l.l.}\right)}=\int \mathrm{d}t\mathrm{d}^{3}x\,N\sqrt{h}\,\mathcal{L}^{\left(\mathrm{l.l.}\right)}\left(h_{ij},K_{ij};R_{ij},\nabla_{i};t\right).\label{S_llap}
	\end{equation}
In (\ref{S_llap}), the spatial derivative $\nabla_{i}$ is restricted to act on the spatial curvature $R_{ij}$ only. 

This kind of action was also discussed within the framework of minimally modified gravity \cite{Lin:2017oow}, in which a condition was derived in order to eliminate the scalar DoF.
In terms of the formalism in the present work, the condition can be written as 
	\begin{widetext}
	\begin{equation}
	\int\mathrm{d}^{3}x\,\sqrt{h}\,\nabla_{m}\left(\frac{1}{\sqrt{h}}\frac{\tilde{\delta}\bar{\mathcal{L}}_{B}^{\mathrm{(M.M.)}}\left[\alpha\sqrt{h}\right]}{\tilde{\delta}R_{kl}}\right)\nabla_{n}\left(\mathcal{A}_{kl}^{ij,mn}B_{ij}\beta\right)-\left(\alpha\leftrightarrow\beta\right)\approx0, \label{mm_2nd_cdt}
	\end{equation}
	\end{widetext}
where the $\alpha$ and $\beta$ are the test functions, $\mathcal{A}_{kl}^{ij,mn}$
is defined in (\ref{A}), and
	\begin{equation}
	\bar{\mathcal{L}}_{B}^{\mathrm{(l.l.)}}\left[\alpha\sqrt{h}\right]\equiv\int\mathrm{d}^{3}x\,\alpha\sqrt{h}\,\bar{\mathcal{L}}_{B}^{\mathrm{(l.l.)}}.
	\end{equation}
In (\ref{mm_2nd_cdt}), $\tilde{\delta}/\tilde{\delta}R_{ij}$ is the functional derivative with respect to $R_{ij}$ while treating $\left(R_{ij},h_{ij},\alpha,\beta\right)$ as independent variables. 
It is interesting to note that the condition (\ref{mm_2nd_cdt}) is nothing but a special case of the second TTDoF condition (\ref{2nd_cdt}), which can be obtained by substituting the action (\ref{S_llap}) into (\ref{2nd_cdt}).
On the other hand, the first TTDoF condition (\ref{1st-cdt}) is automatically satisfied for the action (\ref{S_llap}).

\subsection{The quadratic action}
\label{subsec:The-quadratic-case}

As an illustrating example, in this section we show how the formalism developed in this work applies to a simple action 
	\begin{equation}
	S^{\mathrm{\left(quad\right)}}=\int \mathrm{d}t\mathrm{d}^3x\,N\sqrt{h}\left(b_{1}K^{2}+b_{2}K_{ij}K^{ij}+\mathcal{V} \right),\label{S^quad}
	\end{equation}
with 
	\begin{equation}
	\mathcal{V}\equiv d_{1}+d_{2}R.
	\end{equation}
where the coefficients $b_{1},b_{2}$ etc. are general functions of $t$ and $N$.
Generally, the action (\ref{S^quad}) propagates 3 degrees of freedom, of which two are tensorial and one is the scalar DoF.
We shall show below that as long as the coefficients are required to satisfy the first and the second TTDoF conditions (\ref{1st-cdt}) and (\ref{2nd_cdt}), the scalar DoF will be suppressed. 

According to (\ref{S_B}) and (\ref{eq:S-1}), the action (\ref{S^quad}) can be equivalently written as
	\begin{equation}
	S^{\left(\mathrm{quad}\right)}=S_{B}^{\left(\mathrm{quad}\right)}+\int \mathrm{d}t\mathrm{d}^{3}x\frac{\delta S_{B}^{\mathrm{(quad)}}}{\delta B_{ij}}\left(K_{ij}-B_{ij}\right),
	\end{equation}
where
	\begin{equation}
	S_{B}^{\left(\mathrm{quad}\right)} = \int \mathrm{d}t\mathrm{d}^{3}x\,N\sqrt{h}\left(\mathcal{G}_{\left(2\right)}^{i_{1}j_{1}i_{2}j_{2}}B_{i_{1}j_{1}}B_{i_{2}j_{2}}+\mathcal{V}\right),\label{S_B^(quad)}
	\end{equation}
with
	\begin{equation}
	\mathcal{G}_{\left(2\right)}^{i_{1}j_{1}i_{2}j_{2}}\equiv b_{1}h^{i_{1}j_{1}}h^{i_{2}j_{2}}+b_{2}h^{i_{1}(i_{2}}h^{j_{2})j_{1}}.
	\end{equation}
It is easy to show that
	\begin{equation}
	\frac{\delta S_{B}^{\mathrm{(quad)}}}{\delta B_{ij}}=N\sqrt{h}2\mathcal{G}_{\left(2\right)}^{i_{1}j_{1}ij}B_{i_{1}j_{1}}.
	\end{equation}
	
One may substitute (\ref{S_B^(quad)}) into (\ref{1st-cdt}) and (\ref{2nd_cdt}) to derive the conditions for eliminating the scalar DoF.
Since the Lagrangian belongs to the case discussed in Sec. \ref{subsec:The-extended-cuscuton}, it is more convenient to make use of (\ref{ec_1st_cdt}) and (\ref{ec_2nd_cdt}) directly. 
The second order derivatives are evaluated to be
	\begin{equation}
	\frac{\partial^{2}\left(N\mathcal{L}_{B}^{\mathrm{(quad)}}\right)}{\partial N^{2}}=\frac{\partial^{2}N\mathcal{G}_{\left(2\right)}^{i_{1}j_{1}i_{2}j_{2}}}{\partial N^{2}}B_{i_{1}j_{1}}B_{i_{2}j_{2}}+\frac{\partial^{2}\left(N\mathcal{V}\right)}{\partial N^{2}}, \label{L_B/NN}
	\end{equation}
	\begin{equation}
	\frac{\partial^{2}\mathcal{L}_{B}^{\left(\mathrm{quad}\right)}}{\partial N\partial B_{ij}}=2\frac{\partial\mathcal{G}_{\left(2\right)}^{i_{1}j_{1}ij}}{\partial N}B_{i_{1}j_{1}},\label{L_B/NB}
	\end{equation}
	\begin{equation}
	\frac{\partial^{2}\mathcal{L}_{B}^{\left(\mathrm{quad}\right)}}{\partial B_{kl}\partial R_{mn}}=0,\qquad \frac{\partial^{2}\left(N\mathcal{L}_{B}^{\left(\mathrm{quad}\right)}\right)}{\partial N\partial R_{mn}}=\frac{\partial\left(Nd_{2}\right)}{\partial N}h^{mn}.\label{L_B/BR}
	\end{equation}
The ``inverse'' defined in (\ref{inverse}) is given by
	\begin{equation}
	\mathcal{G}_{i^{\prime}j^{\prime},k^{\prime}l^{\prime}}\left(\vec{x}^{\prime},\vec{y}^{\prime}\right)\equiv\frac{1}{N\sqrt{h}}\mathcal{G}_{i^{\prime}j^{\prime},k^{\prime}l^{\prime}}\delta^{3}\left(\vec{x}^{\prime}-\vec{y}^{\prime}\right),\label{green's_qud-1}
	\end{equation}
where we define
\begin{equation}
\mathcal{G}_{i^{\prime}j^{\prime},k^{\prime}l^{\prime}}\equiv\frac{1}{2}\left(-\frac{b_{1}}{b_{2}\left(3b_{1}+b_{2}\right)}h_{i^{\prime}j^{\prime}}h_{k^{\prime}l^{\prime}}+\frac{1}{b_{2}}h_{i^{\prime}(k^{\prime}}h_{l^{\prime})j^{\prime}}\right),\label{inverse_qud}
\end{equation}
for $b_{2}\neq 0$ and $3b_{1}+b_{2}\neq 0$.

Plugging (\ref{L_B/NN}), (\ref{L_B/NB}) and (\ref{green's_qud-1}) into (\ref{ec_1st_cdt}), the first TTDoF condition for the action (\ref{S^quad}) is shown to be
	\begin{widetext}
	\begin{equation}
	\left(\frac{\partial^{2}\left(N\mathcal{G}_{\left(2\right)}^{i_{1}j_{1}i_{2}j_{2}}\right)}{\partial N^{2}}-4\frac{\partial\mathcal{G}_{\left(2\right)}^{i_{1}j_{1}ij}}{\partial N}N\mathcal{G}_{ij,kl}\frac{\partial\mathcal{G}_{\left(2\right)}^{i_{2}j_{2}kl}}{\partial N}\right)B_{i_{1}j_{1}}B_{i_{2}j_{2}}+\frac{\partial^{2}\left(N\mathcal{V}\right)}{\partial N^{2}}=0.\label{1st_cdt_qud}
	\end{equation}
	\end{widetext}
Generally, it is sufficient that (\ref{1st_cdt_qud})  only holds ``weakly'', that is when the constraints are taken into account.
Here for our purpose, we will assume (\ref{1st_cdt_qud}) hold ``strongly'', that is without using any constraint equations.
Of course, this also implies that we are looking for a particular solution for the coefficients $b_{1}, b_{2}$, etc..
By expanding (\ref{1st_cdt_qud}) explicitly, we arrive at a set of differential equations for the coefficients $b_{1}, b_{2}$, etc.:
	\begin{equation}
	\left(Nd_{1}\right)^{\prime\prime}=0,\qquad\left(Nd_{2}\right)^{\prime\prime}=0,\label{qud_d1_d2}
	\end{equation}
	\begin{equation}
	2b_{2}b_{2}^{\prime}-2Nb_{2}^{\prime2}+Nb_{2}b_{2}^{\prime\prime}=0,\label{qud_b2}
	\end{equation}
	\begin{eqnarray}
	 &  & 2b_{2}b_{1}^{\prime}\left(b_{2}-3Nb_{1}^{\prime}-2Nb_{2}^{\prime}\right)+2b_{1}\left(3b_{2}b_{1}^{\prime}+Nb_{2}^{\prime2}\right)\nonumber \\
	 &  & +Nb_{2}\left(3b_{1}+b_{2}\right)b_{1}^{\prime\prime}=0,\label{qud_b1}
	\end{eqnarray}
where in the above a prime ``$'$'' denotes derivative with respect to $N$.
The solutions to (\ref{qud_d1_d2})-(\ref{qud_b1}) can be found easily, 
	\begin{equation}
	b_{1}=-\frac{N}{3}\left(\frac{\beta_{3}\left(t\right)}{N+\beta_{1}\left(t\right)}+\frac{\beta_{4}\left(t\right)}{N+\beta_{2}}\right),\label{b_1}
	\end{equation}
	\begin{equation}
	b_{2}=\frac{\beta_{4}\left(t\right)N}{\beta_{2}\left(t\right)+N},\label{b_2}
	\end{equation}
	\begin{equation}
	d_{1}=\rho_{1}\left(t\right)+\frac{1}{N}\rho_{3}\left(t\right),\label{d_1}
	\end{equation}
	\begin{equation}
	d_{2}=\rho_{2}\left(t\right)+\frac{1}{N}\rho_{4}\left(t\right),\label{d_2}
	\end{equation}
where the coefficients $\beta_{i}$'s and $\rho_{i}$'s are general functions of $t$ only.
Note in order to recover GR, we assume $\rho_{2}\neq 0$. 

The next step is to solve the second TTDoF condition (\ref{ec_2nd_cdt}), which will put further constraints on the form of the coefficients.
By substituting (\ref{d_1}) and (\ref{b_1}) into (\ref{ec_2nd_cdt}), the second TTDoF condition is shown to be
	\begin{widetext}
	\begin{equation}
	-\rho_{2}\nabla_{q}\left(\frac{-b_{2}\left(3b_{1}+b_{2}-2Nb_{1}^{\prime}\right)+N\left(b_{1}+b_{2}\right)b_{2}^{\prime}}{b_{2}\left(3b_{1}+b_{2}\right)}Bh^{pq}+\frac{b_{2}-Nb_{2}^{\prime}}{b_{2}}B^{pq}\right)\approx0.\label{2nd_cdt_qua}
	\end{equation}
	\end{widetext}
Generally, the solution to (\ref{2nd_cdt_qua}) is complicated if we treat it as a ``strong'' equality. 
Fortunately, it is sufficient that the second TTDoF condition (\ref{2nd_cdt_qua}) only holds weakly.
Comparing with the momentum constraint, which in our case takes the form
	\begin{equation}
	C^{p}\approx-2\sqrt{h}\nabla_{q}\left(b_{1}Bh^{pq}+b_{2}B^{pq}\right)\approx 0,
	\end{equation}
it is interesting that (\ref{2nd_cdt_qua}) takes the same form as that of the momentum constraint.
We thus find a particular solution to (\ref{2nd_cdt_qua}), which satisfies\footnote{Generally, the left and the right-hand-sides of (\ref{b1_sol}) and (\ref{b2_sol}) can be proportional to each other with time-dependent coefficients. Here we simply choose the coefficient to unity.}
	\begin{equation}
	b_{1}=\frac{-b_{2}\left(3b_{1}+b_{2}-2Nb_{1}^{\prime}\right)+N\left(b_{1}+b_{2}\right)b_{2}^{\prime}}{b_{2}\left(3b_{1}+b_{2}\right)}, \label{b1_sol}
	\end{equation}
	\begin{equation}
	b_{2}=\frac{b_{2}-Nb_{2}^{\prime}}{b_{2}}. \label{b2_sol}
	\end{equation}
Together with (\ref{b_1}) and (\ref{b_2}), finally we can solve
	\begin{equation}
	b_{1}=-\frac{1}{3}\left(\frac{2N}{\beta_{1}\left(t\right)+N}+\frac{N}{\beta_{2}\left(t\right)+N}\right),
	\end{equation}
and
	\begin{equation}
	b_{2}=\frac{N}{\beta_{2}\left(t\right)+N}.
	\end{equation}
	
To summarize, for the action in the form of (\ref{S^quad}), we have found a particular solution to the first and the second TTDoF conditions (\ref{ec_1st_cdt}) and (\ref{ec_2nd_cdt}). The corresponding action takes the form
	\begin{eqnarray}
	S^{\mathrm{(quad)}} & = & \int\mathrm{d}t\mathrm{d}^{3}x\,N\sqrt{h}\bigg[\frac{N}{\beta_{2}+N}K^{ij}K_{ij}\nonumber \\
	&  & \quad-\frac{1}{3}\left(\frac{2N}{\beta_{1}+N}+\frac{N}{\beta_{2}+N}\right)K^{2}\nonumber \\
	&  & \quad+\rho_{1}+\rho_{2}R+\frac{1}{N}\left(\rho_{3}+\rho_{4}R\right)\bigg],\label{TTDoF_S_quad}
	\end{eqnarray}
where the coefficients $\beta_{i}$'s and $\rho_{i}$'s are general functions of $t$. 
The action (\ref{TTDoF_S_quad}) describes a class of theories which propagates two tensorial DoFs  without the scalar DoF. 
Note the GR is recovered form (\ref{TTDoF_S_quad}) by setting $\beta_{1}=\beta_{2}=\rho_{3}=\rho_{4}=0$, $\rho_{1}=\text{const}$ and $\rho_{2}=1$. 
The cuscuton theory \cite{Afshordi:2006ad} can be obtained by setting $\beta_{1}=\beta_{2}=\rho_{4}=0$ and $\rho_{2}=1$.

\section{Conclusion \label{sec:Conclusion}}

In this work, we investigated the conditions for a class of spatially covariant gravity theories in order to have two tensorial DoFs only. 
We start from the framework proposed in \cite{Gao:2014soa}, which has been proved to have two tensorial DoFs and one scalar DoF in general \cite{Gao:2014fra,Gao:2018znj,Gao:2018izs}.
By performing a detailed Hamiltonian analysis, we derived the two necessary and sufficient conditions (\ref{1st-cdt}) and (\ref{2nd_cdt}) to eliminate the scalar DoF.
We dub the conditions (\ref{1st-cdt}) and (\ref{2nd_cdt}) as the first and the second TTDoF conditions, respectively. 
The two conditions restrict the form of the Lagrangian in (\ref{S_SCG I}). 
Once the Lagrangian satisfies the two TTDoF conditions, there are only two tensorial DoFs propagating in the theory.

The first TTDoF condition (\ref{1st-cdt}) is derived by requiring the $\tilde{\pi}$ sector in the Dirac matrix is degenerate. 
From (\ref{S_xpl}), the first TTDoF condition (\ref{1st-cdt}) implies that the $\left\{ N,K_{ij}\right\}$-sector in the original Lagrangian is degenerate. One of the main findings in this work is that, the first TTDoF condition is not sufficient to remove the scalar DoF completely. As a result, a second condition is required.

Mathematically, the second TTDoF condition (\ref{2nd_cdt}) is derived by requiring the $C^{\prime}$ sector in the Dirac matrix is degenerate.
Physically, if only the first TTDoF condition is satisfied, there will be odd number of second class constraints and thus the dimension of phase space will be odd at each spacetime point. 
This happens generally in Lorentz breaking theories, in which the primary constraint due to the degeneracy does not necessarily yields a secondary constraint. In fact, this is also similar to the case in \cite{Gao:2018znj}, where two conditions are needed to fully eliminate a DoF.
On the other hand, this is different from the generally covariant theories, in which a primary constraint is always associated with a secondary constraint.

We also compared our results with previous studies.
One class of Lorentz breaking scalar-tensor theories with two DoFs was studied as the generalization of the cuscuton theory \cite{Iyonaga:2018vnu}.
It is interesting that the necessary condition found in \cite{Iyonaga:2018vnu} is nothing but a special case the first TTDoF condition (\ref{ec_1st_cdt}) in our work.
Note our analysis also implies that the second TTDoF condition (\ref{ec_2nd_cdt_test_fct}) is also needed in general.
Another class of Lorentz breaking gravity theories with two tensorial DoFs dubbed as the ``minimally modified gravity''  was proposed in \cite{Lin:2017oow}, in which the Lagrangian is linear in the lapse function.
It is also interesting that the condition derived in \cite{Lin:2017oow} is a special case of the second TTDoF condition (\ref{2nd_cdt}) in our work, while the first TTDoF condition is automatically satisfied.

In order to illustrate how our formalism works, we considered a simple example in which the Lagrangian (\ref{S^quad}) is quadratic in the extrinsic curvature.
After solving the two TTDoF conditions (\ref{ec_1st_cdt}) and (\ref{ec_2nd_cdt}) explicitly, we find one special solution (\ref{TTDoF_S_quad}), which propagates two tensorial DoFs generally and includes GR and the original cuscuton theory \cite{Afshordi:2006ad} as special cases.

Comments are in order. Firstly, if (\ref{tertiary_cst}) is not a tertiary constraint, the constraints $\tilde{\pi}\approx 0$
and $\tilde{C}\approx 0$ must be of the first class as in GR. 
The general covariance has been broken to spatial diffeomorphism in our theory, which corresponds to the
first class constraints $\pi_{i}\approx0_{i}$ and $C_{i}\approx0_{i}$.
The arising of additional first class constraints $\tilde{\pi}\approx 0$
and $\tilde{C}\approx 0$ indicates that there might be an enhanced gauge symmetry in the theory, although which may be different from that of GR.
It is thus interesting to clarify this issue. 
Secondly, as being shown in \cite{Gao:2019lpz} for theories with velocity of the lapse function $\dot{N}$, some of the theories with $\dot{N}$ satisfying the conditions for eliminating the unwanted DoF would be obtained by field transformations from theories which satisfy the conditions trivially (i.e., without $\dot{N}$).
Thus it would be interesting to examine if the spatially covariant gravity theories satisfying the two TTDoF conditions can be related to GR by field transformations. 
As a final remark, the scalar DoF eliminated in this work actually behaves similar to the so-called instantaneous mode discussed in \cite{DeFelice:2018mkq}, that is, it disappears in the spatially covariant gravity formulation but will reappear when the general covariance is apparently recovered.
Although it is argued that such an instantaneous mode is safe by choosing appropriate boundary conditions, it is still worth looking at the behaviour of such a mode in our framework in details.

\acknowledgments

This work was supported by the National Youth Thousand Talents Program of China (No. 71000-41180003), by the Natural Science Foundation of China under the grant No. 11975020 and by the SYSU start-up funding.

\appendix

\section{Derivation of (\ref{S_xpl})}
\label{sec:The-first-TTDoF}

In this appendix, we show the explicit derivation of the first TTDoF condition (\ref{1st-cdt}) and (\ref{S_xpl}). 
Instead of calculating the Poisson bracket $\left[\tilde{\pi}\left(\vec{x}\right),C^{\prime}\left(\vec{y}\right)\right]$ in (\ref{PB_pitld_Cprime}) directly, it is more convenient to introduce a test function $Y(\vec{x})$ and evaluate
	\begin{eqnarray}
	0 & \approx & \left[\int\mathrm{d}^{3}x\,Y(\vec{x})\tilde{\pi}(\vec{x}),C'(\vec{y})\right]\nonumber \\
	& \approx & \left[\int\mathrm{d}^{3}x\left(p^{ij}(\vec{x})X_{ij}(\vec{x})+\pi(\vec{x})Y(\vec{x})\right),\left[H_{\mathrm{C}},\pi(\vec{y})\right]\right],\qquad \label{PB_pit_Cp_xpl}
	\end{eqnarray}
where we have used the definitions of $C'$ and $\tilde{\pi}$ in (\ref{C^prime}) and (\ref{pi^tilde}), as well as the fact that
	\begin{equation}
		\left[\tilde{\pi}(\vec{x}),\tilde{\pi}^{kl}(\vec{y})\right]= 0.
	\end{equation}
In (\ref{PB_pit_Cp_xpl}), $X_{ij}$ is determined by $Y$ through (\ref{X_kl}).

By evaluating the Poisson brackets in (\ref{PB_pit_Cp_xpl}) explicitly, we get
	\begin{widetext}
	\begin{eqnarray}
	0 & \approx & \int\mathrm{d}^{3}x\left[p^{ij}\left(\vec{x}\right),\left[H_{\mathrm{C}},\pi\left(\vec{y}\right)\right]\right]X_{ij}\left(\vec{x}\right)+\int\mathrm{d}^{3}x\left[\pi\left(\vec{x}\right),\left[H_{\mathrm{C}},\pi\left(\vec{y}\right)\right]\right]Y\left(\vec{x}\right)\nonumber \\
	& = & -\int\mathrm{d}^{3}x\left(\left[\pi\left(\vec{y}\right),\left[p^{ij}\left(\vec{x}\right),H_{\mathrm{C}}\right]\right]+\left[H_{\mathrm{C}},\left[\pi\left(\vec{y}\right),p^{ij}\left(\vec{x}\right)\right]\right]\right)X_{ij}\left(\vec{x}\right)+\int\mathrm{d}^{3}x\left[\pi\left(\vec{x}\right),\left[H_{\mathrm{C}},\pi\left(\vec{y}\right)\right]\right]Y\left(\vec{x}\right)\nonumber \\
	& = & -\int\mathrm{d}^{3}x\left(\left[\pi\left(\vec{y}\right),-2N\tilde{\pi}^{ij}\left(\vec{x}\right)\right]\right)X_{ij}\left(\vec{x}\right)+\int\mathrm{d}^{3}x\frac{\delta^{2}S_{B}}{\delta N\left(\vec{x}\right)\delta N\left(\vec{y}\right)}Y\left(\vec{x}\right)\nonumber \\
	& \approx & \int\mathrm{d}^{3}x\frac{\delta^{2}S_{B}}{\delta N\left(\vec{x}\right)\delta N\left(\vec{y}\right)}Y\left(\vec{x}\right)+\int\mathrm{d}^{3}xN\left(\vec{x}\right)\frac{\delta}{\delta N\left(\vec{y}\right)}\left(\frac{1}{N\left(\vec{x}\right)}\frac{\delta S_{B}}{\delta B_{ij}\left(\vec{x}\right)}\right)X_{ij}\left(\vec{x}\right)\nonumber \\
	& \equiv & \int\mathrm{d}^{3}x\,Y\left(\vec{x}\right)\bigg[\frac{\delta^{2}S_{B}}{\delta N\left(\vec{x}\right)\delta N\left(\vec{y}\right)}-\int\mathrm{d}^{3}x'\int\mathrm{d}^{3}y'\,N\left(\vec{x}'\right)\frac{\delta}{\delta N\left(\vec{x}\right)}\left(\frac{1}{N\left(\vec{x}'\right)}\frac{\delta S_{B}}{\delta B_{i^{\prime}j^{\prime}}\left(\vec{x}'\right)}\right)\nonumber \\
	&  & \qquad\times\mathcal{G}_{i^{\prime}j^{\prime},k^{\prime}l^{\prime}}\left(\vec{x}',\vec{y}'\right)N\left(\vec{y}'\right)\frac{\delta}{\delta N\left(\vec{y}\right)}\left(\frac{1}{N\left(\vec{y}'\right)}\frac{\delta S_{B}}{\delta B_{k^{\prime}l^{\prime}}\left(\vec{y}'\right)}\right)\bigg],
	\end{eqnarray}
	\end{widetext}
where we have used some expressions of the Poisson brackets in (\ref{eq:csc_cdt})
and (\ref{X_kl}). It thus immediately follows that (\ref{PB_pitld_Cprime}) implies (\ref{1st-cdt}) and (\ref{S_xpl}).


%

\end{document}